\documentclass[usegraphicx]{nature}
\spacing{1}

\usepackage{graphicx}
\usepackage{lineno}
\usepackage{wasysym}
\usepackage{amssymb}
\usepackage{xcolor}
\usepackage{soul}
\usepackage{amstext}
\usepackage{array}
\usepackage{caption}

\newcolumntype{L}{>{$}l<{$}}
\newcolumntype{C}{>{$}c<{$}}
\newcolumntype{R}{>{$}r<{$}}

\DeclareRobustCommand{\ion}[2]{\textup{#1\,\textsc{\lowercase{#2}}}}

\newcommand{\rcs}{RCSGA 032727$-$132623}
\newcommand\arcsec{\mbox{$^{\prime\prime}$}}%
\newcommand\arcmin{\mbox{$^{\prime}$}}%
\newcommand{\za}{$z_{\rm abs}$}
\newcommand{\zg}{$z_{\rm G1}$}

\newcommand{\mgii}{\ion{Mg}{ii}}
\newcommand{\feii}{\ion{Fe}{ii}}
\newcommand{\mgi}{\ion{Mg}{i}}
\newcommand{\oii}{\ion{O}{ii}}

\newcommand{\kms}{km~s$^{-1}$}

\newcommand*\ga{\aa@centerstack{>}{\sim}}

\newcommand{\mnras}{Mon. Not. R. Astron. Soc.}
\newcommand{\apj}{Astrophys. J.}
\newcommand{\apjs}{Astrophys. J. Suppl. Ser.}
\newcommand{\aj}{Astronom. J.} 
\newcommand{\apjl}{Astroph. J. Letters}
\newcommand{\aap}{Astron. \& Astrophys. }
\newcommand{\pasp}{Publications of the Astronomical Society of the Pacific}
\newcommand{\aaps}{ Astronomy \& Astrophysics Supplement Series}
\newcommand{\araa}{  Annual Review of Astronomy \& Astrophysics}

\title{\noindent 
A clumpy and anisotropic  galaxy halo at {\textsl z=1} from gravitational-arc
tomography   
}

\author{\noindent 
Sebastian Lopez$^{1,*}$, 
Nicolas Tejos$^{2}$, 
C\'edric Ledoux$^{3}$,
L. Felipe Barrientos$^{4}$,  
Keren Sharon$^{5}$,  
Jane R. Rigby$^{6}$,  
Michael D. Gladders$^{7,8}$,  
Matthew B. Bayliss$^{9}$,  
\& 
Ismael Pessa$^{4}$ 
}

\begin{document}

\maketitle

\begin{affiliations}
\begin{footnotesize}
 \item Departamento de Astronom\'ia, Universidad de Chile,
    Casilla 36-D, Santiago, Chile; $^{*}$slopez@das.uchile.cl  % 1
  \item Instituto de F\'isica, Pontificia Universidad Cat\'olica de
    Valpara\'iso, Casilla 4059, Valpara\'iso, Chile % 2 
 \item European Southern Observatory, Alonso de C\'ordova 3107, Casilla 19001,
    Vitacura, 19, Santiago, Chile % 3
  \item Instituto de Astrof\'isica, Pontificia Universidad Cat\'olica de Chile,
    Vicu\~na Mackenna 4860, Santiago, Chile % 4
   \item Department of Astronomy, University of Michigan, 500 Church St., Ann
    Arbor, MI 48109, USA  % 5
  \item Observational Cosmology Lab, NASA Goddard Space Flight Center, 8800
    Greenbelt Rd., Greenbelt, MD 20771, USA % 6
  \item Department of Astronomy \& Astrophysics, University of Chicago, 5640
    S. Ellis Ave., Chicago, IL 60637, USA  % 7
  \item Kavli Institute for Cosmological Physics, University of Chicago, 5640
    S. Ellis Ave., Chicago, IL 60637, USA % 8
\item Kavli Institute for Astrophysics \& Space Research, Massachusetts
  Institute of Technology, 77 Massachusetts Avenue, Cambridge, MA 02139, USA % 9
\end{footnotesize}
\end{affiliations}

\begin{abstract}

Every star-forming galaxy has a halo of metal-enriched gas extending out to at
least 100~kpc\cite{Churchill2000,Chen2010,Nielsen2013}, as revealed by the
absorption lines this gas imprints on the spectra of background
quasars\cite{Chen2016}. However, quasars are sparse and typically probe only
one narrow pencil beam through the intervening galaxy.  Close quasar
pairs\cite{Tytler2009,Martin2010,Hennawi2006} and gravitationally lensed
quasars\cite{Rauch2001,Lopez2007,Chen2014,Rubin2017} have been used to
circumvent this inherently one-dimensional technique, but these objects are
rare and the structure of the circum-galactic medium remains poorly
constrained. As a result, our understanding of the physical processes that drive
the re-cycling of baryons across the lifetime of a galaxy is
limited\cite{Keres2005,Muratov2015}. Here we report integral-field
(tomographic) spectroscopy of an extended background source ---a bright giant
gravitational arc. We can thus coherently map the spatial and kinematic
distribution of \mgii\ absorption ---a standard tracer of enriched gas--- in
an intervening galaxy system at redshift $0.98$ (i.e., $\sim 8$ Gyr ago). Our
gravitational-arc tomography unveils a clumpy medium in which the
absorption-strength decreases with increasing impact parameter, in good
agreement with the statistics towards quasars; furthermore, we find strong
evidence that the gas is not distributed isotropically. Interestingly, we
detect little kinematic variation over a projected area of $\approx 600$
kpc$^2$, with all line-of-sight  velocities confined to within a few tens of
\kms\ of each other. These results suggest that the detected absorption
originates from entrained recycled material, rather than in a galactic
outflow.

\end{abstract}

\begin{figure}
\includegraphics[angle=0,scale=0.445,clip]{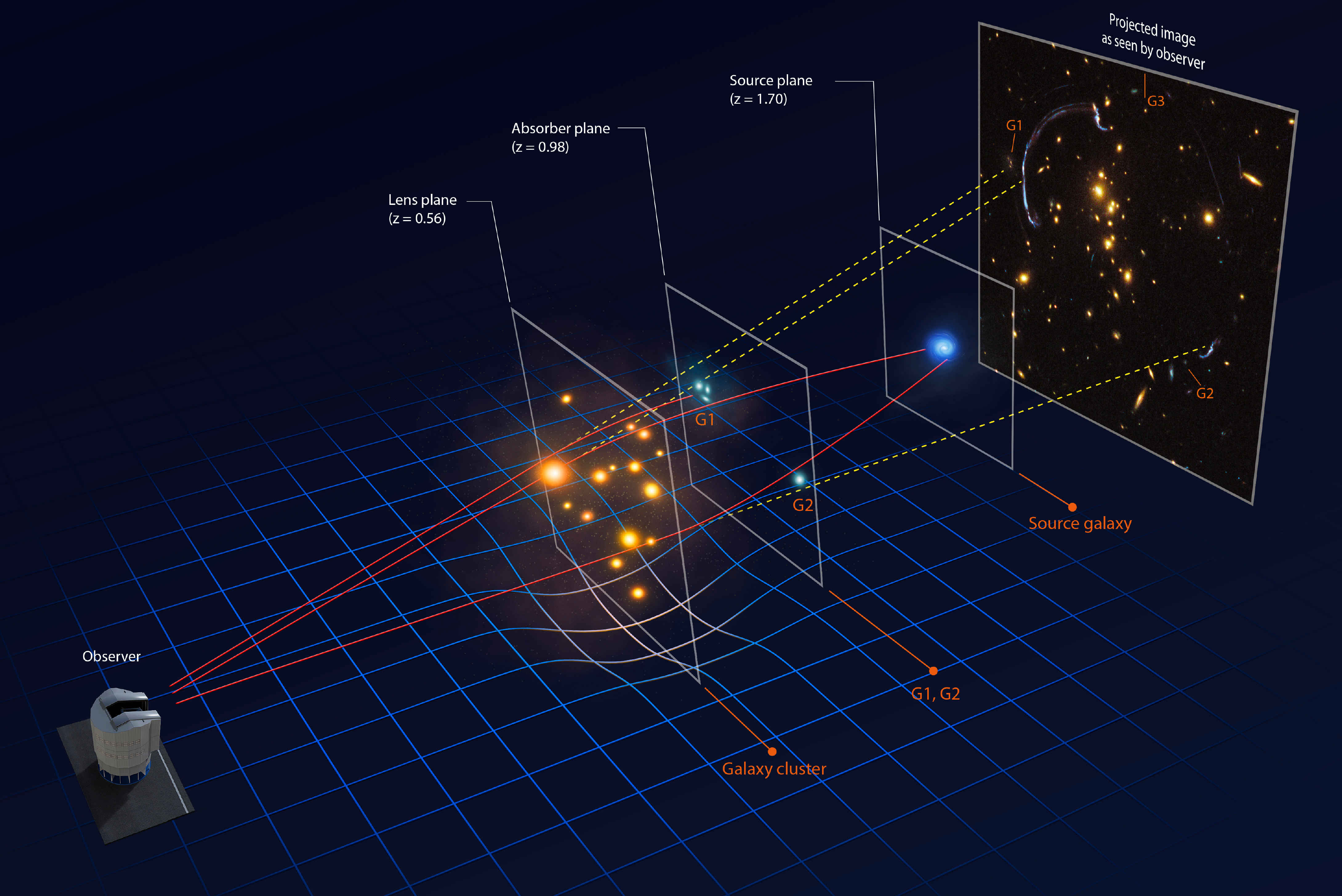}
\caption{{\bf Illustration of the lens geometry of arc and absorber in \rcs.}
  Light from the $z=1.70$ background galaxy ('source plane') is deflected 
    and magnified by an intervening galaxy cluster at $z=0.56$\cite{Wuyts2010}
    ('lens 
    plane'), to 
  form the 
  bright giant arc that is seen in the 'image plane' (right-most panel).  As
  the light crosses the \za$=0.98$ absorber plane, some of it is absorbed by
  \mgii\ in the gas that surrounds an absorbing galaxy lying close in
  projection. Three candidates for such galaxies are detected at this
  redshift, marked 'G1', 'G2' and 'G3' in the image plane. The present work
  deals with G1 and G2 (indicated also in the absorber plane), which probe the
  closest projected distances to the arc. (Image produced by Carlos Polanco).
}
\label{fig_lens_layers}
\end{figure}

We use the Multi Unit Spectroscopic Explorer (MUSE)\cite{Bacon2010} mounted on
the European Southern Observatory Very Large Telescope to observe the
$38$\arcsec\ long gravitational arc \rcs\cite{Wuyts2010}.  This arc
  results from a lensed galaxy at redshift $1.70$, highly magnified and
  stretched by a massive galaxy cluster at redshift $0.56$ 
(Figure~\ref{fig_lens_layers}).  With $g=19.15$, it is among the brightest
known arcs\cite{Dahle2016} and has a high surface brightness across a large
area on the sky\cite{Rigby2011}.  Magellan/MagE spectroscopy\cite{Rigby2017}
of its brightest knot reveals the presence of a strong \mgii\ absorption
system at redshift \za$=0.98$, and we set out to map this absorption along the
entire arc.

\begin{figure}
\includegraphics[angle=0,scale=0.54,clip]{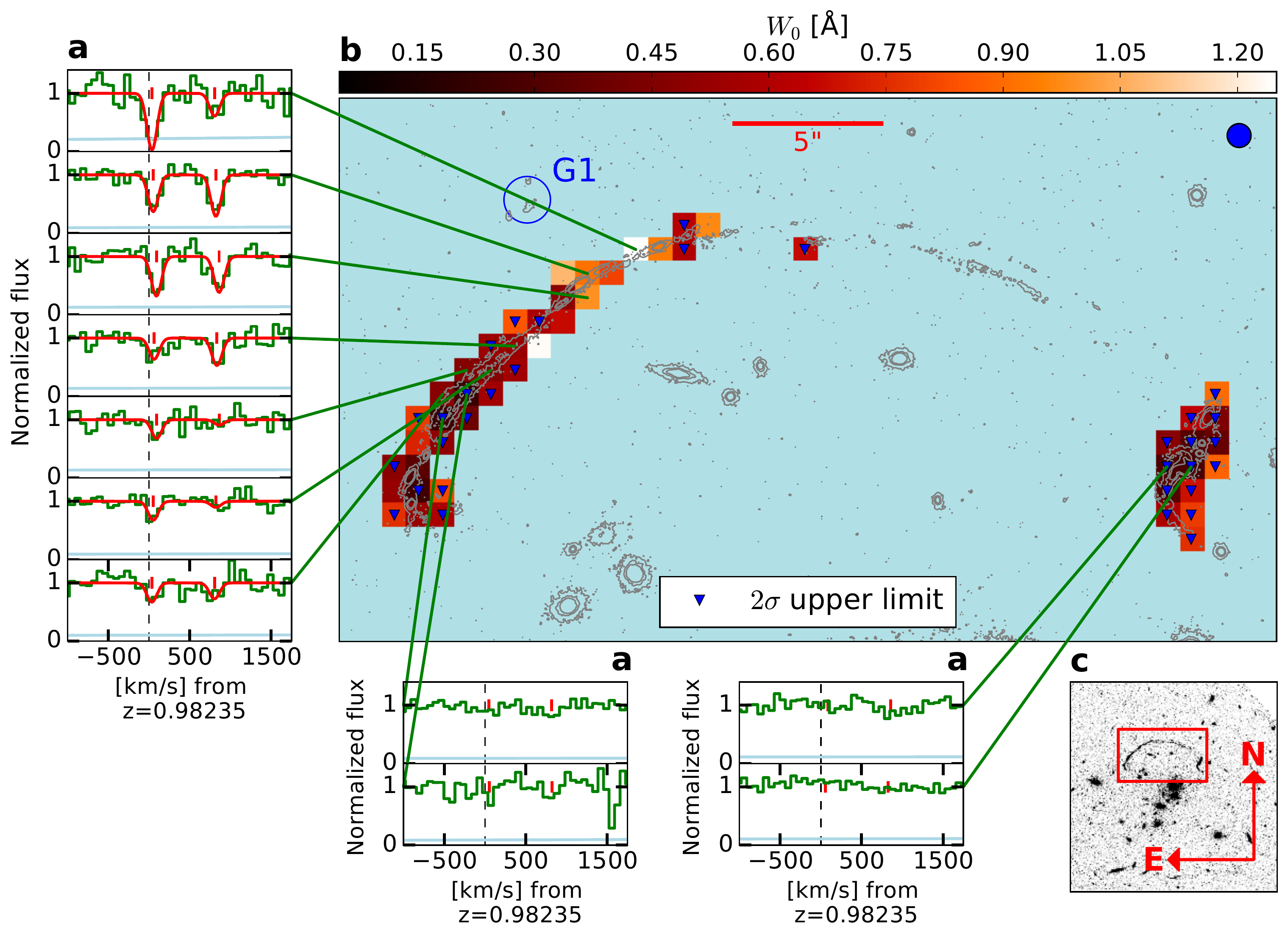}
\caption{{\bf Map of \ion{Mg}{ii} absorption-strengths at $\approx 4$ kpc
    resolution.}  {\bf (a):} 
Representative sample of \ion{Mg}{ii}$\lambda\lambda 2796,2803$ MUSE
  spectra (green histograms) and corresponding Gaussian fits (red lines) in
  velocity space with respect to \zg$=0.98235$.  Panels without fits
  correspond to non-detections. The green lines indicate the corresponding
  positions on the arc.  {\bf (b):} $0.51\arcmin\times0.27$\arcmin\ inset of
  the $1\arcmin\times 1$\arcmin\ MUSE field centered in the north-eastern part
  of the arc.  The color scale indicates \mgii$\lambda 2796$ rest-frame
  absorption-strengths obtained from the Gaussian fits.  A total of 56
  positions were selected to have a continuum signal-to-noise ratio S/N$>3$ at
  the expected \mgii\ lines (50 shown here), out of which 18 have significant
  detections (see Methods).  Non-detections are indicated by blue downward
  arrows in the map. The candidate absorber, G1, is indicated by the blue
  circumference (to the North-East).  For reference, we overlay arc and galaxy
  contours at $840$ nm from {\it Hubble Space Telescope (HST)} data (GO
  program 12267; PI: Rigby).  Each independent spatial element ('spaxel') is
  0.8\arcsec 
  wide, equivalent to 4 MUSE unbinned spaxels and matching the seeing
  (indicated by the blue circle on the top-right).  This grid is defined in
  the image plane; the actual spatial resolution varies across the absorber
  plane from $\approx 4$ kpc, at the eastern side of the arc, to $\approx 2$
  kpc, at the western side.  Likewise, the $5\arcsec$ scalebar corresponds to
  a range of $\approx24$--$12$ kpc in the absorber plane, depending on
  position (for a de-lensed image see Extended Data
  Figure~\ref{fig_lens}).  {\bf (c): } Entire MUSE field-of-view indicating
  the location of the \mgii\ map shown in (b).
\label{fig_rEW_map}
}
\end{figure}

Figure~\ref{fig_rEW_map} shows a MUSE map of \za$=0.98$ \ion{Mg}{ii}
absorption-strength at different positions along \rcs.  The arc positions span
impact parameters ---in the absorber plane; see Methods--- of $\approx
15$--$90$ kpc to the absorbing galaxy system (hereafter ``G1''), indicated by
the blue circle. G1 is our primary candidate absorber (out of three [\oii]
detections at redshift $0.98$, namely G1, G2, and G3; Methods) and deserves
special attention.  {\it HST} continuum images resolve this system into
three irregular galaxies, hereafter G1-A, G1-B and G1-C, all having blue $B-I$
colors.  From the {\it HST} photometry we estimate these galaxies to have low
luminosities, $\lesssim 0.05~{\rm L}^*$, and consequently also low stellar
masses, compared to quasar absorbers\cite{Nielsen2013}.  Such stellar masses
suggest total halo masses of $\sim 10^{11}$ M$_{\rm \astrosun}$, which in turn
define their virial radii to be $R_{\rm vir}\sim90$ kpc (Methods).
Additionally, from the [\oii]$\lambda\lambda 3726,3729$ emission doublets, we
estimate their star-formation rates to amount to $\lesssim 0.2$ M$_{\rm
  \astrosun}$~yr$^{-1}$.  We use the [\oii] velocities to define a
``systemic'' redshift at \zg$=0.98235$.

Four key features are readily evident from Figure~\ref{fig_rEW_map}: (a)
\mgii\ is detected from $\approx 15$ out to $\approx 45$ kpc to the South of
G1 but not detected further than $\approx 80$ kpc to the West (another arc
knot in the South ---not shown in the Figure--- also has sensitive
non-detections; see below); (b) the absorption-strength is not uniform,
indicating a clumpy medium on 4 kpc scales down to our detection limit of
$\approx 0.4$ \AA; (c) the line centroids vary little (within one spectral
pixel, or $\sim50$ \kms) all across the arc; and (d) most of the doublet
ratios appear to be saturated, indicating possible partial covering of the
background source (Methods).

\begin{figure}
\includegraphics[angle=0,scale=0.65,clip]{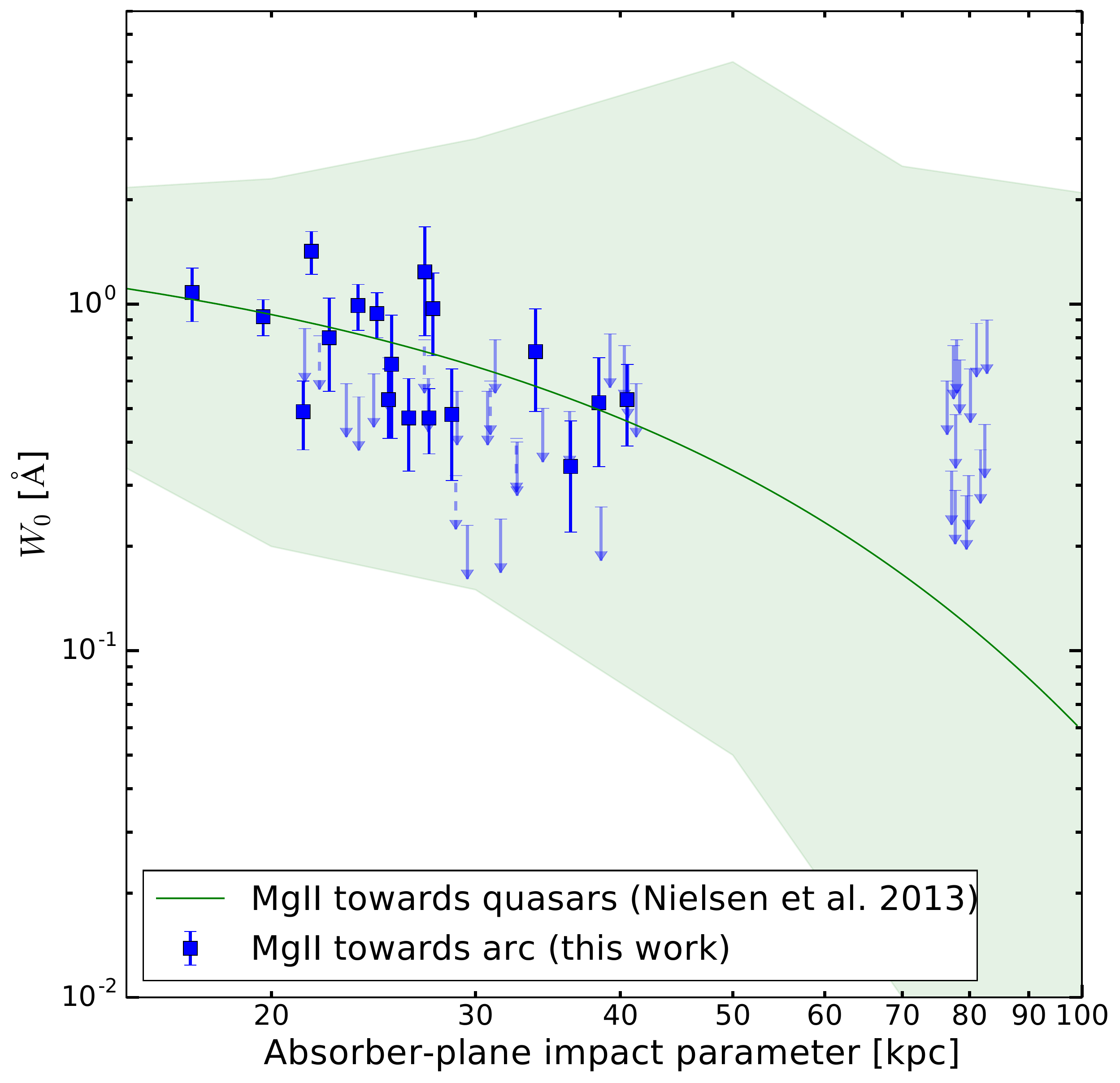}
\caption{{\bf Arc \mgii\ absorption-strength versus impact parameter.}
  Squares correspond to detections and arrows to $2\sigma$ upper limits. 
 Error bars correspond to the
$\pm 1\sigma$ uncertainty in the $W_0$ measurement. All
  impact parameters are measured to G1 (North-East part of the arc), except
  for the six upper limits marked with dashed lines which correspond to G2
  (South).  The zero-points are given by the centers of the blue circles
    in Figure~\ref{fig_rEW_map} and Extended Data Figure~\ref{fig_oii}.  The
  green line is the 
  fit to a sample of $182$ quasar absorption systems\cite{Nielsen2013} and the
  shaded area indicates the full scatter in that sample.  Error bars
    correspond to the $\pm 1\sigma$ uncertainty in the $W_0$
    measurement. Uncertainties in the impact parameters are $\lesssim 
  5$\%.
\label{fig_rEW_ip}
}
\end{figure}

Figure~\ref{fig_rEW_ip} shows that the absorption-strength decreases with
  impact parameter, in  broad agreement with the quasar
statistics\cite{Nielsen2013}, but the scatter is {\it smaller} than in the
quasar data\cite{Chen2010,Nielsen2013}. The latter is likely a consequence of
partial covering, which would skew down the arc measurements (Methods);
however, the heterogeneity of the compiled quasar-galaxy sample may also play
a role\cite{Nielsen2013,Nielsen2013cat}, as the intervening galaxies 
encompass a wide range of masses, luminosities, and orientations. That is to
say, we compare averages over different areas in a single galaxy (probed by
the arc) with an average over many  distinct galaxies (probed by
quasars).

From quasar lens statistics\cite{Rauch2001,Lopez2005}, we know transverse
structure is detected on similar scales as probed by our 4 kpc, seeing-limited
resolution, and below. This indicates that the metals traced by \mgii\ are
concentrated in small clouds we do not resolve here but are spatially
distributed in such a way to produce the gradient we observe. Therefore, the
present data do not probe individual cloud sizes but rather their coherence
length. Some stringent non-detections (Figure~\ref{fig_rEW_ip}) re-enforce
the notion of clumpiness on kpc scales.

\begin{figure}
\includegraphics[angle=0,scale=0.65,clip]{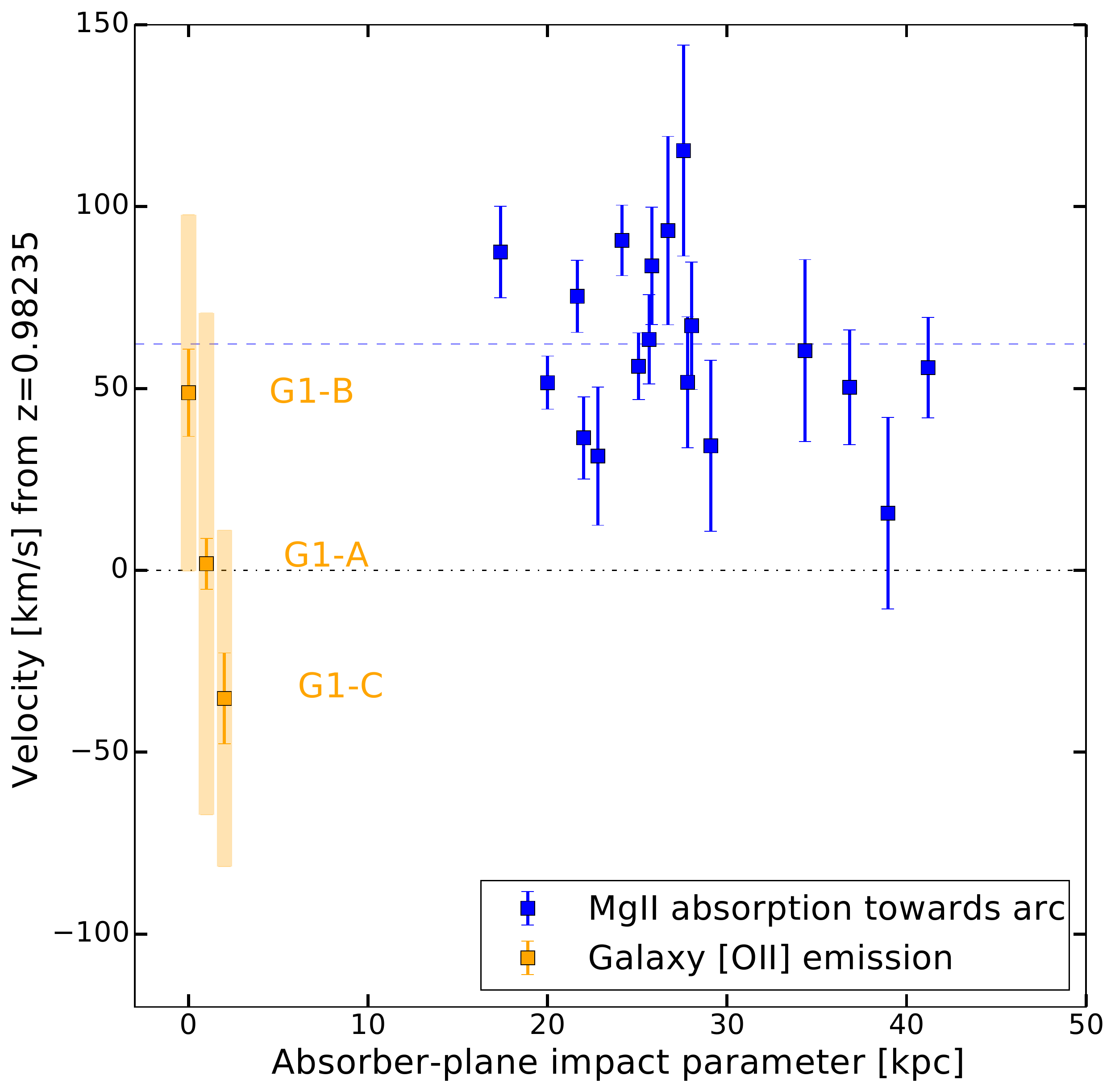}
\caption{{\bf Gas kinematics.} Blue symbols correspond to arc
  \ion{Mg}{ii}$\lambda\lambda 2796,2803$ absorption-line velocities; orange
  symbols to galaxy [\ion{O}{ii}]$\lambda\lambda 3726,3729$ emission line
  velocities (shifted in the x-axis for clarity).  The blue dashed line
  indicates the average absorption velocity. The zero-point velocity
  corresponds to the systemic redshift at \zg$=0.98235$.
  Error bars correspond to the $\pm 1\sigma$ uncertainty in the velocities.
  The envelopes around the emission-line measurements indicate the FWHM
  velocity dispersion derived from the [\oii] fits.  Uncertainties in the
  impact parameters are $\lesssim 5$\%.
\label{fig_vel_ip}
}
\end{figure}

The present tomographic technique allows us to scan the velocity field of the
absorbing gas profusely in a single high-redshift
halo\cite{Chen2014}. Figure~\ref{fig_vel_ip} displays absorption velocities
and emission velocities of G1-A, G1-B and G1-C as a function of impact
parameter. The first outstanding feature in this Figure is that all absorption
velocities lie to the red of \zg, at $+62$ \kms, although none of them
substantially exceeds any of the galaxy velocity dispersions.  Secondly, there is
little variation in absorption velocity  ($\approx24$ \kms) overall
along the arc, even less than within the galaxy system itself, but extending
out to 40 kpc, i.e. 10 times larger distances than the projected separations
between the G1 galaxies!  Along with a clumpy medium revealed by the map of
absorption-strengths, this kinematically {\it quiet} behavior places strong
constraints on the geometry and dynamics of this system.

Assuming saturation, the absorption-strengths are a measure of the velocity
spread of individual clouds\cite{Ellison2006} (not resolved by our data) and
possible partial covering (see Methods). Thus, most of our detections would
correspond to velocity spreads $\lesssim108$ \kms\ along the sightlines, the
equivalent of a $1$ \AA\ absorption line.  Interestingly, we find a lower
scatter in the {\it transverse} direction. Taken together, this would indicate
gas clouds whose internal velocity dispersions dominate over bulk motions. For
the impact parameters and halo mass probed here, these velocities resemble
those determined at higher spectral resolution in low-redshift
systems\cite{Tumlinson2013,Ho2017}, which appear well within the halo escape
velocity and virial radius.

Our observations do not support a spherical
geometry\cite{Steidel1995,Charlton1996}. Indeed, the \mgii\ gas does not seem
to be distributed isotropically around G1; if this were the case, similar
absorption would occur at both sides of the line connecting G1 and its closest
arc position, which we do not see. Instead, there are more non-detections on
the Southern side. G2 (Methods) is also a good example of this situation, as no
\mgii\ is detected in six positions at $\approx 20$--$30$ kpc to sensitive
limits, while 4 are expected if G2 followed the trend shown in
Figure~\ref{fig_rEW_ip} isotropically. Furthermore, assuming the observed
\mgii\ to be related to only one of the G1 galaxies, then only one out of four
galaxies (including G2) presents detectable absorption, which leads to a rough
covering fraction of $\approx 25$\% within 40 kpc.  This interpretation
assumes that the G1 galaxies are distinct objects and not part of a large disk
of gas and dust (Methods).  Thus, in line with quasar-absorber
observations\cite{Bouche2012,Kacprzak2012}, our arc observations strongly
suggest that the absorbing gas is anisotropic, showing wide (possibly
$\sim90^\circ$) opening angles\cite{Bordoloi2014b}.

Our data also allow us to compute gas covering fractions in an alternative
fashion, namely {\it directly} from our ``hits and misses'' statistics around
G1 only (Methods). Assuming anisotropy, we probe here the preferential G1
direction that shows absorption; therefore, our covering fraction estimate
should lie {\it above} that of quasar absorbers, because those surveys probe
random quasar-galaxy orientations. Interestingly enough, our prediction is
fulfilled in comparison with {\it galaxy-selected} quasar
absorbers\cite{Chen2010}, which we regard to be unbiased: the
  covering fraction towards the arc is indeed larger than towards the
  quasars in that sample (Methods). Conversely, a comparison with more
  heterogeneous samples that include {\it \mgii-selected}
  galaxies\cite{Nielsen2013} gives smaller covering fractions towards the arc
  than towards the quasars.  This suggests possible selection biases in the
latter samples, because by construction they favour galaxy orientations
producing absorption\cite{Nielsen2013cat}.

At {\it low redshift} there is observational
evidence\cite{Ho2017,Kacprzak2017} that the circum-galactic medium of
star-forming galaxies is driven by the interplay between major-axis entraining
gas (pristine or re-cycled) and enriched outflows aligned with the minor
axis. Although this picture is also consistent with most recent simulations
and $\Lambda$CDM predictions\cite{Muratov2017}, it remains yet to be
confirmed, particularly at high redshift. Here we deal with dwarf galaxies at
redshift one, still able to eject metals\cite{Keres2005} out to one
  virial radius, beyond which the metal flux is expected to have decreased
  markedly\cite{Muratov2017}. The present \mgii\ detections occur well within
  $R_{\rm vir}$ and therefore could be originating from any of the G1 galaxies
  (but given the relatively quiet velocity field most probably from only one
  of them).  In addition, the gas is metal-enriched (including \mgii\ and \feii; see
  Methods) and patchy (perhaps revealing a multi-phase medium), which suggests
  re-cycled gas either outflowing from or bound to one of the individual G1
  halos.  The outflow scenario requires outflow
  speeds\cite{Martin2010,Bouche2012} in excess of the galaxy-gas velocity
  offsets observed here (Figure~\ref{fig_vel_ip}).  This suggests that
  the \mgii\ is more 
  likely correlated with entrained gas in the extended halo of G1-B,
   (closer in velocity) or 
  even G1-A, in which case the absorption velocity offset would resemble the
  one-sided velocity offsets seen at low redshift\cite{Tumlinson2013,Ho2017}.

The novel gravitational-arc tomography presented here appears to
probe the gaseous extension of a galaxy halo {\it in formation}, beyond
$\approx20$ kpc scales, which might be a remnant of past gravitational
interactions forming tidal debris and gaseous streams infalling back into the
overall G1 potential well. Unfortunately, the arc-galaxy configuration under
study 
does not cover lower impact parameters for testing this hypothesis, but future
objects may permit more conclusive detections.

\subsection{References}
\bibliographystyle{naturemag}

\begin{addendum}
 \item This work has benefitted from discussions with Alain Smette, Nicole
   Nielsen and Glenn Kacprzak.  S.L. thanks the European Southern Observatory
   Scientific Visitor Selection Committee for supporting a research stay at
   the ESO headquarters in Santiago, where part of this work was done.
   S.L. has been supported by FONDECYT grant number 1140838. This work has
   also been partially supported by PFB-06 CATA. 
N.T. acknowledges support from {\it CONICYT PAI/82140055. }

 \item [Author contributions]{
S.L. conceived and led the project. S.L. and N.T. wrote the MUSE
telescope-time proposal and designed the observations.  L.F.B. and
N.T. prepared the remote observations and L.F.B. reduced the MUSE data. S.L.,
N.T. and C.L. analysed the data, performed simulations, and devised original
ways to produce and interprete the results. S.L. wrote the main
codes. N.T. and I.P. performed the blind survey of galaxies in the field of
view. K.S. performed the lens  model and L.F.B. supervised the design of
Fig. 1.  M.B.B. and L.F.B., 
performed the photometric characterization of the absorbing galaxies, and
S.L., C.L. and N.T. the analysis of their spectra.  Ancillary data from MagE
and HST were provided by J.R.R. and M.D.G. S.L. wrote the manuscript and
crafted the rest of the figures, with contribution from N.T. All co-authors
provided critical feedback and helped shape the manuscript.  }
 \item[Competing Interests] The authors declare that they have no
competing financial interests.
 \item[Correspondence] Correspondence and requests for materials
should be addressed to\\ S. Lopez~(email: slopez@das.uchile.cl).
\end{addendum}

\clearpage
\begin{methods}

\section{Observations and data reduction.}

The data were obtained with the MUSE integral-field
spectrograph\cite{Bacon2010} mounted on the ESO Very Large Telescope. The
1\arcmin\ field-of-view is sampled with $349\times352$ $0.2$\arcsec\ wide
spaxels. MUSE uses image slicers combined with an array of 24 identical
spectrographs that provide nearly $100\,000$ spectra simultaneously. Our setup
provided a wavelength range from $4\,650$ to $9\,300$ \AA\ at a resolving
power $R=\lambda/\Delta \lambda$ ranging from $2\,000$ to $4\,000$. Each
spectral bin is $1.25$ \AA\ wide. The observations were carried out in
'service mode' during dark time, with thin cirrus or clear-sky conditions,
airmass below 1.8, and seeing better than 0.8\arcsec.  We obtained a total of
21 exposures of $700$s on-target each. The exposures were taken within
``Observing Blocks'' of 4 exposures each. We applied a small dithering and
$90^{\circ}$ rotations between exposures to reject cosmic rays and minimize
patterns of the slicers on the processed combined cube.  We reduced all the
observations using the MUSE pipeline recipe v1.6.4 and ESO reflex v2.8.5. We
analyzed the quality of the individual exposures by measuring the seeing and
transparency in the ``white'' images. Due to changes in the weather conditions
during integrations some of the exposures come from aborted observing blocks
(but are otherwise $700$s long).  We discarded 5 of them for having guiding
errors, poor PSF, or obvious transparency problems. The remaining 16 exposures
were combined into one final science datacube. The total on-target time was
therefore $3.1$ hours.  The sky subtraction was improved on this cube using
the Zurich Atmospheric Purge (ZAP) algorithm\cite{Soto2016}. The wavelength
solution, corrected for the motion of the Earth around the Sun and converted to
vacuum, was checked satisfactorily at the position of the arc's brightest knot
against our MagE spectrum\cite{Rigby2017}. The spectral resolution at the
\ion{Mg}{ii} absorption lines is ${\rm FWHM} = 2.7$ \AA\, corresponding to
$\approx 140$ \kms\ (or $R=2\,100$).

In this work we adopt a flat $\Lambda$CDM cosmology with $\Omega_\Lambda =
0.7$, $\Omega_m = 0.3$, and $H_0 = 70$ km~s$^{-1}$~Mpc$^{-1}$.

\section{Spaxel combination.}

Unlike quasars, the present source is (a) extended and (b) inhomogeneous in
flux given the resolved background galaxy and the inhomogeneous lensing 
magnifications (Extended Data Figure~\ref{fig_binning}). Therefore, the spectrum extraction
must proceed using an ad-hoc flux combination to assure independent flux
measurements while maximizing the final S/N per spectral pixel.  To first
approximation we tried polygons of constant S/N\cite{Cappellari2003} but
decided not to use them given that widely different areas complicate the
interpretation of absorption against an extended source. Instead, we perform a
'minimum-size' binning based on a fixed grid that was defined by the spatial
sampling. Since the seeing profile is sampled with four $0.2$\arcsec\ spaxels
(FWHM) we use $4\times4$ binning. This means that the center of each binned
spaxel lies at least one seeing FWHM from each other. Offsetting the grid by
$\pm 1$ unbinned spaxel shows that the choice of zero-point does not affect
the results of the present analysis.

\section{Spectral extraction.}

To extract individual 1-D spectra for each binned spaxel (or ``positions'') we
applied weights given by $w_{ij}=f_{ij}/v_{ij}$, where $f_{ij}$ and $v_{ij}$
are respectively the calibrated object flux and variance of spaxel
$(i,j)$. This weight is a modified version of that used to optimally combine
spectra of different S/N\cite{Robertson1986}; it assumes a low detector
read-out noise, in which case $v_{ij}$ approximates to the total number of
counts. We note that this gives higher weights to brighter spaxels and
may introduce a complicated bias towards highly magnified spaxels.  Since we
were interested in a small spectral range we defined a 'sub-cube' containing
wavelengths between $5\,100$ and $6\,000$ \AA. This range includes the
\ion{Mg}{ii} doublet and the transitions \feii$\lambda2382$, 
  \feii$\lambda2600$, and \mgi$\lambda2852$ at \za$=0.98$. Since the array of
variances is noisy in the spectral direction as well, to avoid introducing
noise carried by the spectral weights, we chose a weight integrated in the
spectral direction in a small featureless region bluewards of the \ion{Mg}{ii}
lines (the red side is compromised by a sky emission line residual). Also,
this weight does not penalize the absorption troughs.

\section{Spaxel selection and Gaussian fits}

To search for \mgii\ we first created a S/N map by selecting binned spaxels on
top of the arc and having S/N $>3$ to the blue of the expected \ion{Mg}{ii}
absorption. This selected a total of $56$ positions, corresponding to a total
projected area surveyed of $\approx 600$ kpc$^2$. At each selected position a
spectrum was extracted and an automatic Gaussian fit applied to the spectral
region corresponding to \zg$=0.98235$. The \ion{Mg}{ii}$\lambda\lambda
2796,2803$ doublet was fitted with two Gaussians having a tied wavelength
ratio, free doublet ratio and fixed line width (corresponding to the
instrument spectral resolution, FWHM = 2.7 \AA).  Each fit provides a
rest-frame absorption-strength (rest-frame equivalent width or hereafter
``$W_0$'') and a redshift (radial velocity, hereafter ``$v$''), along with
their statistical errors (Extended Data Tables~\ref{table_absorption}
and~\ref{table_absorption_G2}).  Fixing the line width provides more
  robust fits, as expected, avoiding false positives in low S/N regions. It
  also assumes the instrumental profile dominates the line profile.  This may
  not hold for all the positions, in which case we estimate a systematic error
  of $\approx 0.05$ \AA\ would be introduced in $W_0$ (not included in the
  tables).

In addition to the fits, synthetic line-profiles were created
for comparison with the data.  When a fit failed or the significance was below
3$\sigma$, a $2\sigma$ upper limit was calculated using the formula
$\sigma_{W_0}(1+z)={\rm FWHM}/\langle{\rm S/N}\rangle$, where $\sigma_{W_0}$
is the expected rest-frame $1\sigma$ error in the $W_0$ measurement and
$\langle{\rm S/N}\rangle$ the average continuum S/N near the line.  The 
  procedure delivers a total of 18 significant \mgii\ detections, all of them
  to the North-East of the arc. Finally, to create the absorption-strength
and velocity maps the fit $W_0$ and $v$ values (or the $W_0$ upper limits)
were recorded in images having the same spatial sampling as the S/N map.

We attempted simultaneous fits to \feii$\lambda2382$ and \mgi$\lambda2852$ 
tied to the  \mgii\ redshifts (unfortunately, the 
\feii$\lambda2600$ 
transition is blended with the
[\ion{C}{iii}]$\lambda\lambda 1907,1909$ emission line doublet
  from the source galaxy\cite{Rigby2015} and therefore it could not be 
  fitted). We detect significant \feii\ at only three positions along the arc 
  (Extended Data Table~\ref{table_absorption_FeII}).  These positions also show the
  strongest {\it and } most significant \mgii\ absorption.  The
  corresponding \feii/\mgii\ strength ratios lie in the rather narrow range
  $0.5$--$0.7$ and conform to the quasar statistics of {\it very strong}
  ($W_0>1$ \AA) \mgii\ systems\cite{Rodriguez-Hidalgo2012}.  Non-detections do
  not constrain these ratios at other positions of the arc due to either too
  weak \mgii\ or too low S/N. Although the present 
  \feii\ data are limited, the similar ratios tentatively  hint to homogeneous
  enrichment along the arc. One of these positions  shows a (marginal) \mgi\ detection.

The absorption signal in the maps, though significant, is expected to be weak
and directly affected by the inhomogeneous S/N. To rule out possible artifacts
due to reduction or analysis effects, we conducted mock tests by simulating a
flat $W_0$ distribution in a cube having the same S/N per spaxel as the actual
data. The outcome of these tests shows that our fitting procedure recovers a
true signal in nearly 100\% of the (binned) spaxels with S/N$>~3$, which 
justifies our S/N cutoff.

\section{Galaxies at $z=0.98$}

We searched systematically for galaxies in the MUSE cube near $z=0.98$. The
search included continuum sources and emission-line galaxies. We detected a
total of 3 [\ion{O}{ii}] sources, which we refer to as G1, G2 and G3
(Extended Data Figure~\ref{fig_oii}). These form a triangle with sides 42\arcsec, 47\arcsec
and 64\arcsec-long, or $d\approx 231$, $259$ and $341$ kpc in the absorber
plane.  G1 is resolved into 3 galaxies in the {\it HST} continuum images,
which are barely resolved by MUSE. We refer to them as G1-A, G1-B and
G1-C. From Gaussian fits to the [\oii]$\lambda\lambda 3726,3729$ doublet in
the MUSE spectra we obtained redshifts, line velocity dispersions, [\oii]
luminosities, and star-formation rates for the five galaxies.  Note that
  the deconvolved velocity dispersions are subject to systematics, given the
  modest MUSE resolution. Assuming virial equilibrium, the velocities of
  G1, 
G2 and G3 lead to a virial radius $R_{\rm vir}< d$ and thus we do not consider
them to be bound gravitationally given their projected separations.  Instead,
we deem them as 3 independent systems that lie in the same large-scale structure
at $z=0.98$.

G1-A, G1-B and G1-C represent our best candidate absorbing galaxy system, 
  due to their proximity to the place where the arc absorption occurs. The
  close proximity of G1-A, G1-B and G1-C may cast doubts on whether these
  indeed are distinct galaxies, as opposed to a single disk where dust
  obscuration would mimic the presence of different objects. The fact that
  they are resolved also in the rest-frame $I$-band (F160W)
  supports the existence of distinct galaxies.

We obtained the photometry from {\it HST} images in the F606W, F814W, F098M,
F125W and F160W bands. We used SExtractor\cite{Bertin1996} in dual mode using
the detections in the F160W band as reference to obtain AB magnitudes in a
0.8\arcsec-diameter aperture. $(B-I)$ rest-frame colors were computed from
F814W and F160W given these filters are the closest in effective
wavelength. We used a local Scd galaxy spectral template\cite{Coleman1980}
that represents well the color of these galaxies, to correct for any mismatch
in the effective pass-bands. Small extinction corrections\cite{Schlegel1998}
were also applied.  Using the multi-band photometry and a SED fitting
code\cite{Moustakas2013} we estimated luminosities, stellar masses and
star-formation rates.  These quantities are subject to large uncertainties due
to the use of just 5 passbands.  We also computed star-formation rates for
each galaxy from the [\oii] line fluxes, integrated over 16 unbinned
spaxels. These line fluxes are subject to extinction and therefore they must
be treated as lower limits only.  Using Kennicutt\'{}s
prescription\cite{Kennicutt1998} the emission-line luminosities translate into
star-formation rates that are broadly consistent with those obtained from the
SED fitting. These values were corrected by magnification $\mu$ using our lens
model. From the stellar masses we estimate dark-matter halo masses, $M_h$,
using the prediction from abundance-matching\cite{Moster2010}. We then
determine the corresponding virial radius using the relation $R_{\rm
  vir}=(3M_h/200\rho_c4\pi)^{1/3}$, where $\rho_c$ is the critical density of
the Universe at
$z=0.98$.  The galaxy data are summarized in Extended Data
Table~\ref{table_galaxies}.

\section{Lens model}
 
We base our lens model on a previous lensing analysis of this
cluster\cite{Sharon2012}.  We improve the previous lens model with new lensing
constraints, including three new spectroscopic redshifts that we measured from
the MUSE data. We obtain a spectroscopic redshift for the radial arc S7a/S7b
in Sharon et al. (2012), $z_{spec}=2.82624$. We identify two new sets of
lensed galaxies, at $z_{spec}=2.7$ and $z_{spec}=5.2$. 
Including more spectroscopic redshift
constraints substantially reduces the lens model uncertainties and improves the
accuracy\cite{Sharon2012,Johnson2016}. 
The lens model is computed using the public software
{\tt{Lenstool}}\cite{Jullo2007}, which uses a Markov Chain Monte Carlo (MCMC)
process 
to explore the parameter space. 
The lens model results in a computed projected mass density distribution 
  in 
the lens plane, magnification maps for any given redshift, deflection fields,
and their uncertainties. The deflection matrices are calculated using the lens equation,
 $\vec\beta = \vec\theta - d_{ls}/d_s \vec\alpha(\vec\theta)$, where
 $\vec\beta$ is the source position at $z_{source}$, $\vec\theta$ is the
 observed position, $d_{ls}$ and $d_s $ are the distances from the lens to the
 source and from the observer to the source, respectively, and
 $\vec\alpha(\vec\theta)$ is the deflection angle at the observed position. We
 note that any plane behind the lens can be considered as source plane,
 and 
 this equation applies as well to the absorber plane.

To assess the completeness of our search for [\oii] in emission, we
scanned the magnification map near \mgii\ looking for regions with much lower
magnification than on top of G1 and G2. We found none, indicating that these
galaxies are not sitting on particularly highly-magnified regions;
consequently we can be confident that our absorber candidates are robust and
no other galaxies (of similar brightness but non-magnified) got missed from
our search.

\section{Spatial resolution and impact parameters}

 We use the deflection matrices to de-lens the coordinates of our binned
 spaxels into the absorber plane and to calculate impact parameters.
 Extended Data Figure~\ref{fig_lens} shows a de-lensed image of the arc
 projected against 
 the image plane.  Due to the inhomogeneous lensing deflection, when the
 spaxels are traced from the image plane to the absorber plane the shape of
 the binned spaxels changes in the absorber plane, although there is no
 overlap between them. Assuming the light rays do not intercept each other
 after being absorbed, our signal should probe independent areas of the
 absorber. We discuss below on the unequal spaxel areas.

From Extended Data Figure~\ref{fig_lens} it also becomes evident that although
the angular 
resolution is constant across the image plane, the actual spatial resolution 
---defined in the absorber plane--- is not.  To define an ad-hoc 'spatial
resolution' we simply take the square root of the area of each de-lensed
spaxel. We find that this number ranges from $\approx 4$ kpc at the East
side of the arc, to $\approx 2$ kpc at the western side.

To calculate impact parameters we multiply the de-lensed angular separations
between spaxels and G1 by the scale at $z=0.98$ given by the adopted
cosmology, i.e., 
$7.97$ kpc/$\arcsec$. 
From a large
set of MCMC realizations, we estimate the statistical $1\sigma$ uncertainty
associated with the angular separations to be $\lesssim 2$\%. Including model
systematics\cite{Johnson2016}, impact parameters should be precise at the 
$\lesssim 5$\% level for the assumed cosmology.

\section{Partial covering}

The background source is likely to be more extended than the typical size of
the absorbing clouds, leading to possible partial
covering\cite{Bergeron2017}. To test this effect on our signal we performed a
second run of automatic \mgii\ fits on a $8\times8$ spaxel
map. Extended Data Figure~\ref{fig_covering} shows the cumulative
distributions of $W_0$ and 
$v$ for the $4\times4$ and $8\times8$ binnings. The stronger binning indeed
skews the $W_0$ distribution to lower values.  This is probably due to
averaging arc light without \mgii\ absorption; although given the arc
geometry, at this spaxel size, 1.6\arcsec, we expect also some sky
contamination (not so in the $4\times4$ binning). Conversely, {\it the
  velocity distribution remains unaffected}.  We conclude that (a) the chosen
$4\times4$ binning is our best option above the seeing limitation; (b) we
cannot exclude a level of partial covering in our $W_0$ sample, although such
an effect is not affecting the absorption velocities.

Particularly relevant to interpreting our results, the physical areas covered
by the binned spaxels are unequal in the absorber plane
(Extended Data Figure~\ref{fig_lens}). These areas are of the order of $10$
kpc$^2$, i.e., 
at least $10^7$ times larger than those probed by the $\lesssim 1$-pc quasar
beams. Clearly, our arc measurements sample an {\it average} signal at each
position. Therefore, measurements at different positions are comparable with
each other and, to a great extent, also {\it independent} of the spaxel
area. The only effect of having uneven areas in the absorber plane should
be on the intrinsic scatter of each measurement, with smaller spaxels having
more scatter. Obviously one cannot measure this intrinsic scatter, but we
consider its effect negligible given that the goal is to make a comparison
with measurements obtained along the much narrower quasar beams.

\section{Gas covering fractions}
To calculate gas covering fractions we compute the fraction of positions
having 
positive detections above a given $W_0$ cutoff, $W_{\rm cut}$, in a given
impact parameter ($D$) range. Non-detections are accounted for by considering
only (2$\sigma$) upper limits below $W_{\rm cut}$.  We select $W_{\rm
  cut}$ and ranges of $D$ to enable comparisons with two quasar-absorber
statistics.  

To compare with the survey presented by Chen and collaborators\cite{Chen2010}
we assume an average galaxy magnitude $\langle M_B \rangle = -19.0$ and use
$W_{\rm cut}=0.5$ \AA.  We find  covering fractions of $100\%$ (2/2; in 
  the range $16<D<20$ kpc), $80\%\ (4/5; 20<D<25$ kpc), and $38\%\ (6/16;
  25<D<39$ kpc). The quasar statistics for low-luminosity galaxies
  gives\cite{Chen2010} $60$\%, $20$\% and $0$\%, respectively for the three
  $D$ ranges. These figures suggest that the intervening gas covering fraction
  is larger towards this arc than towards the quasars in that sample.

On the other hand, to compare with the survey presented by Nielsen and
collaborators\cite{Nielsen2013} we use two cutoffs: $W_{\rm cut}=0.6$ \AA\ and
$W_{\rm cut}=1.0$ \AA.  For $W_{\rm cut}=0.6$ \AA\ we find covering fractions
of $67\%\ (6/9; 0<D<25$ kpc), and $20\%\ (4/20; 25<D<50$ kpc).  The quasar
statistics for low-luminosity galaxies ($0.07<L_B/L_B^*<0.94$)
gives\cite{Nielsen2013} $71$\% and $48$\%, respectively. For $W_{\rm cut}=1.0$
\AA\ we find covering fractions of $18\%\ (2/11)$ and $4 \%\ (1/26)$,
respectively for the same $D$ ranges. The quasar statistics gives $29$\% and
$24$\%. Thus, for both $W_0$ cutoffs, the covering fractions towards the arc
are {\it smaller} than towards the quasars in this sample.

Finally, note that all these comparisons are subject to uncertainties because
we do not cover exactly the same published impact-parameter ranges.

\end{methods}

\subsection{References for Methods}

\begin{addendum}
\item[Code availability]{ 
The present analysis is based on
custom Python routines, some of which use the MUSE Python Data Analysis
Framework\cite{Bacon2016} and the open-source plotting package for Python
APLpy\cite{Robitaille2012}. We have opted not to make our routines available
because these are described in detail in the paper.
} 
\item[Data availability] 
The observations discussed in this paper were made with European Southern
Observatory (ESO) Telescopes at the La Silla Paranal Observatory under
programme ID 098A.0459(A). The corresponding data are available on the ESO
archive  at http://archive.eso.org/cms.html.  
\end{addendum}

\renewcommand{\thefigure}{\arabic{figure}}
\renewcommand{\figurename}{Extended Data Figure}
\setcounter{figure}{0}

\clearpage

\begin{figure}
\includegraphics[angle=0,scale=0.58,clip]{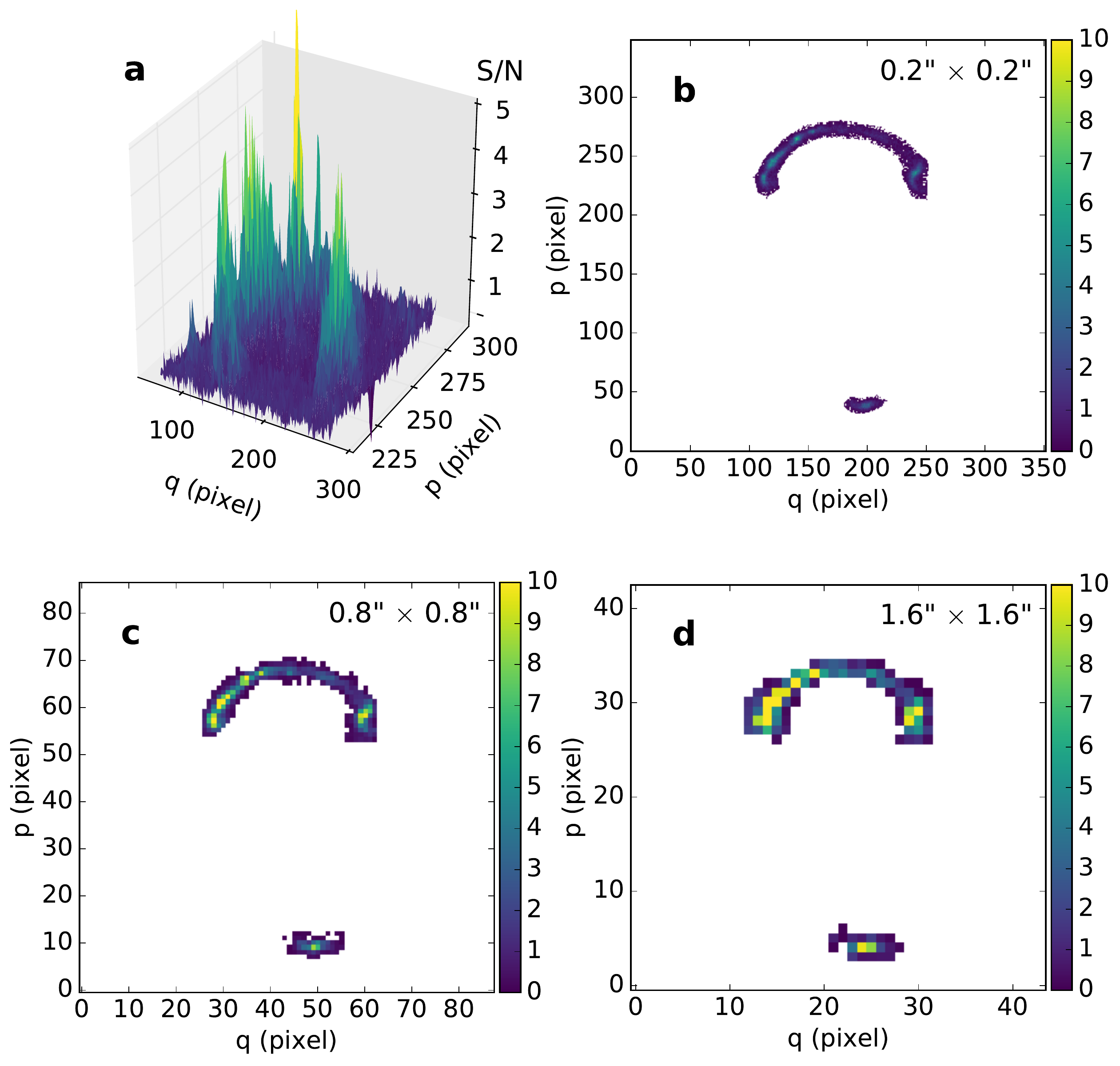}
\caption{{\bf S/N versus binning.} {\bf (a):} 3-D representation of the
  S/N at the position of \mgii\ absorption in the unbinned data.  {\bf (b)---(d):} Same as (a) but in 2-D and for different
  binnings. The size of each binned spaxel is indicated in
  arc-seconds. The 
  color scale is the same for all three panels. Note the expected increase in S/N with
  binning size.
\label{fig_binning}
}
\end{figure}

\begin{figure}
\includegraphics[angle=0,scale=0.58,clip]{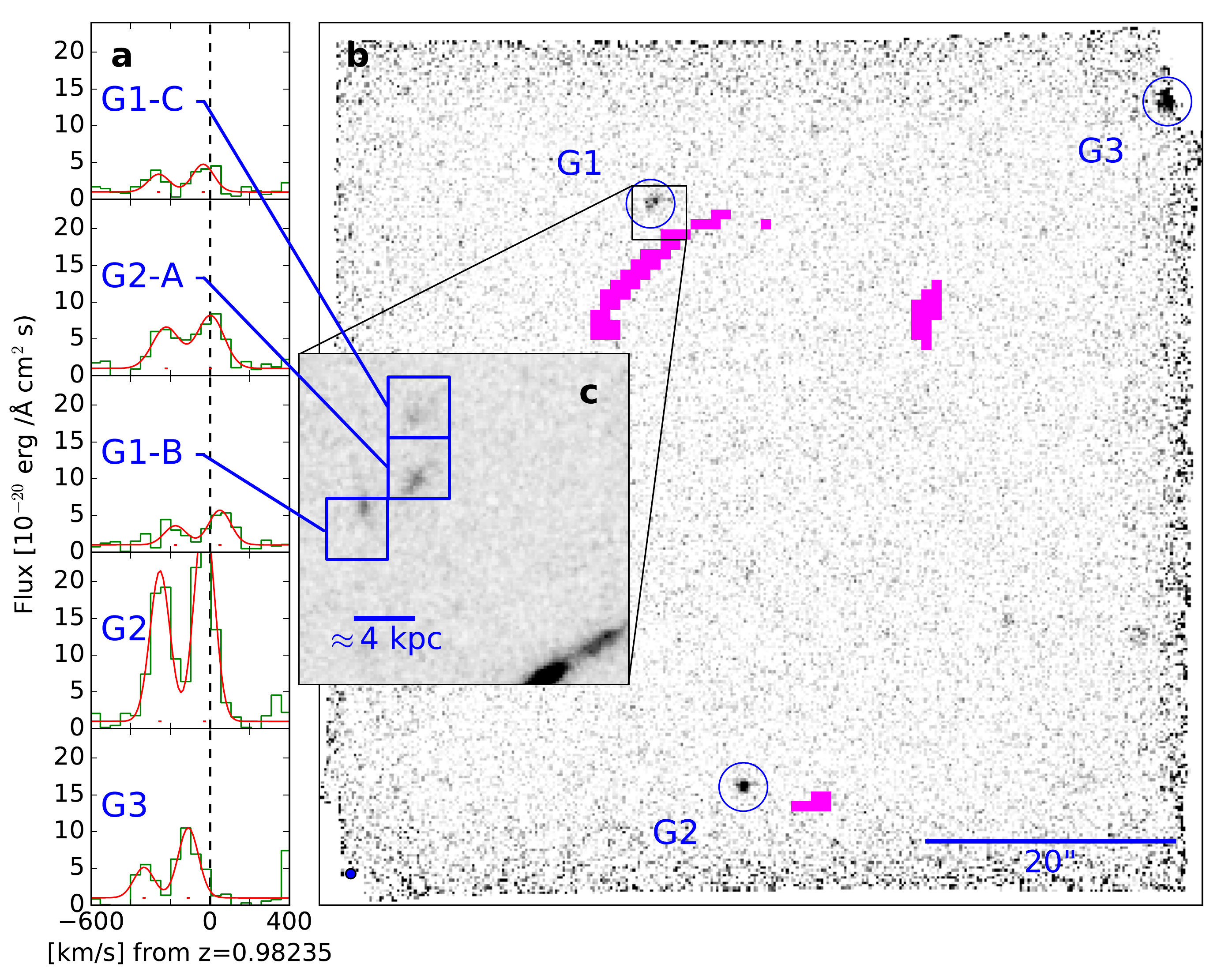}
\caption{ {\bf Emission-line galaxies at $z=0.98$.}  {\bf (a):} Gaussian fits
  to the [\oii]$\lambda\lambda 3726,3729$ doublets in the MUSE spectra of G1,
  G2, and G3, the three [\oii] sources found by our systematic search.  The
  MUSE spatial resolution barely resolves G1 into three [\oii] clumps (G1-A,
  G1-B, and G1-C), which cluster around \zg$=0.98235$ and have a velocity
  dispersion of $35$ \kms.  {\bf (b):} MUSE image of \rcs\ centered on [\oii]
  emission at $z=0.98$. The magenta squares indicate the binned spaxels used
  to map the \mgii\ absorption against the arc.  {\bf (c):} {\it HST}/WFC3
  F814W image zooming into the G1 system. The blue squares indicate the MUSE
  regions used to extract the [\oii] spectra.  The scale corresponds to the
  region close to G1 in the absorber plane.
\label{fig_oii}
}
\end{figure}

\begin{figure}
\includegraphics[angle=0,scale=0.85,clip]{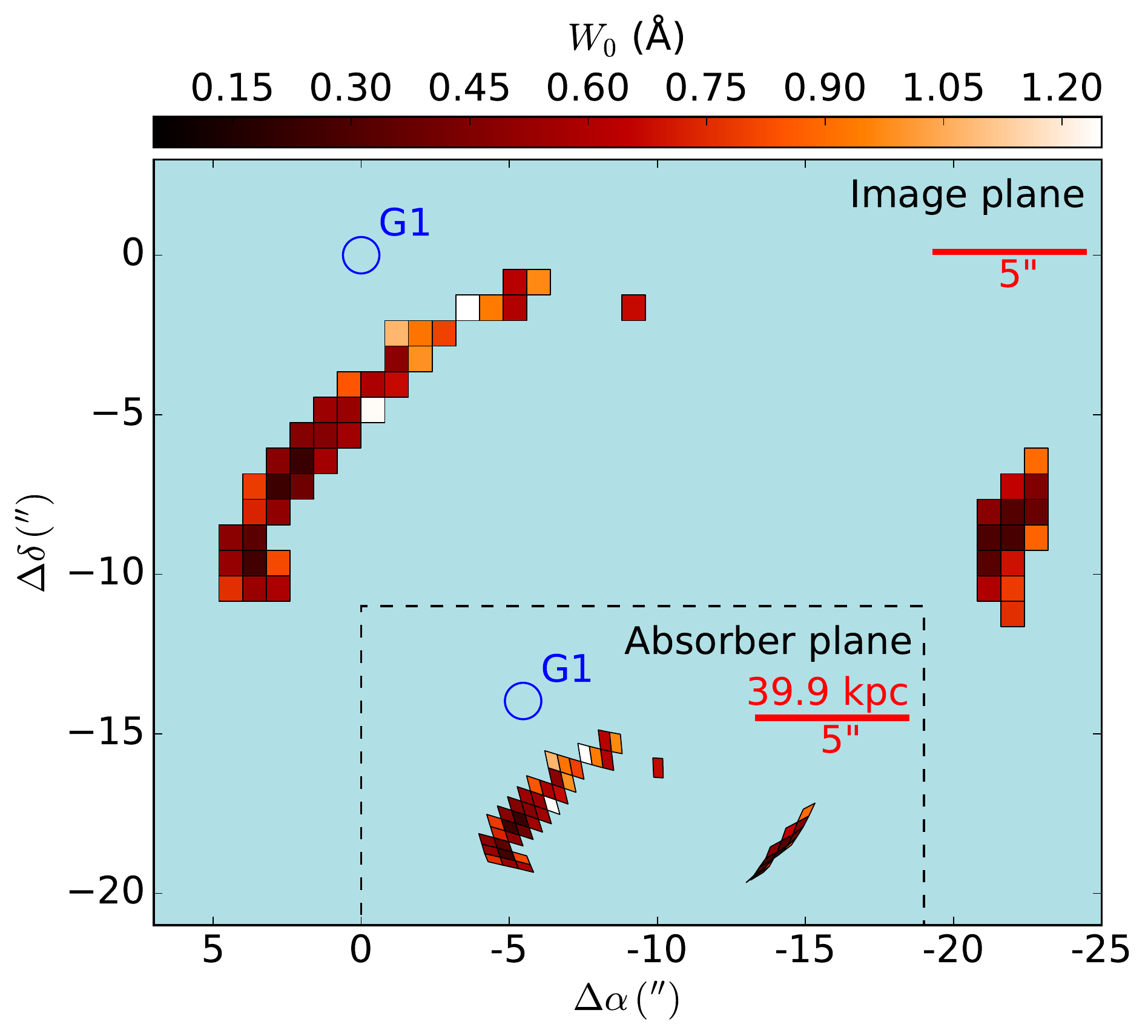}
\caption{{\bf Projection of absorber plane against image plane.} In the
  absorber plane (dashed-line rectangle) the spaxel configuration appears
  shrunk and the de-lensed spatial elements have different shapes and areas
  across the absorber plane.  After de-lensing, the scale in the absorber
  plane is given by the adopted cosmology, i.e., $5\arcsec =39.85$ kpc at
  $z=0.98$.  The impact parameter used in this work is defined as the
  projected physical distance between a given position and G1 on this plane.
  For reference, a $5\arcsec$ scale bar is shown in the image plane too.
  Coordinates are in arc-seconds relative to the position of G1 in the image
  plane.
\label{fig_lens}
}
\end{figure}

\begin{figure}
\includegraphics[angle=0,scale=0.58,clip]{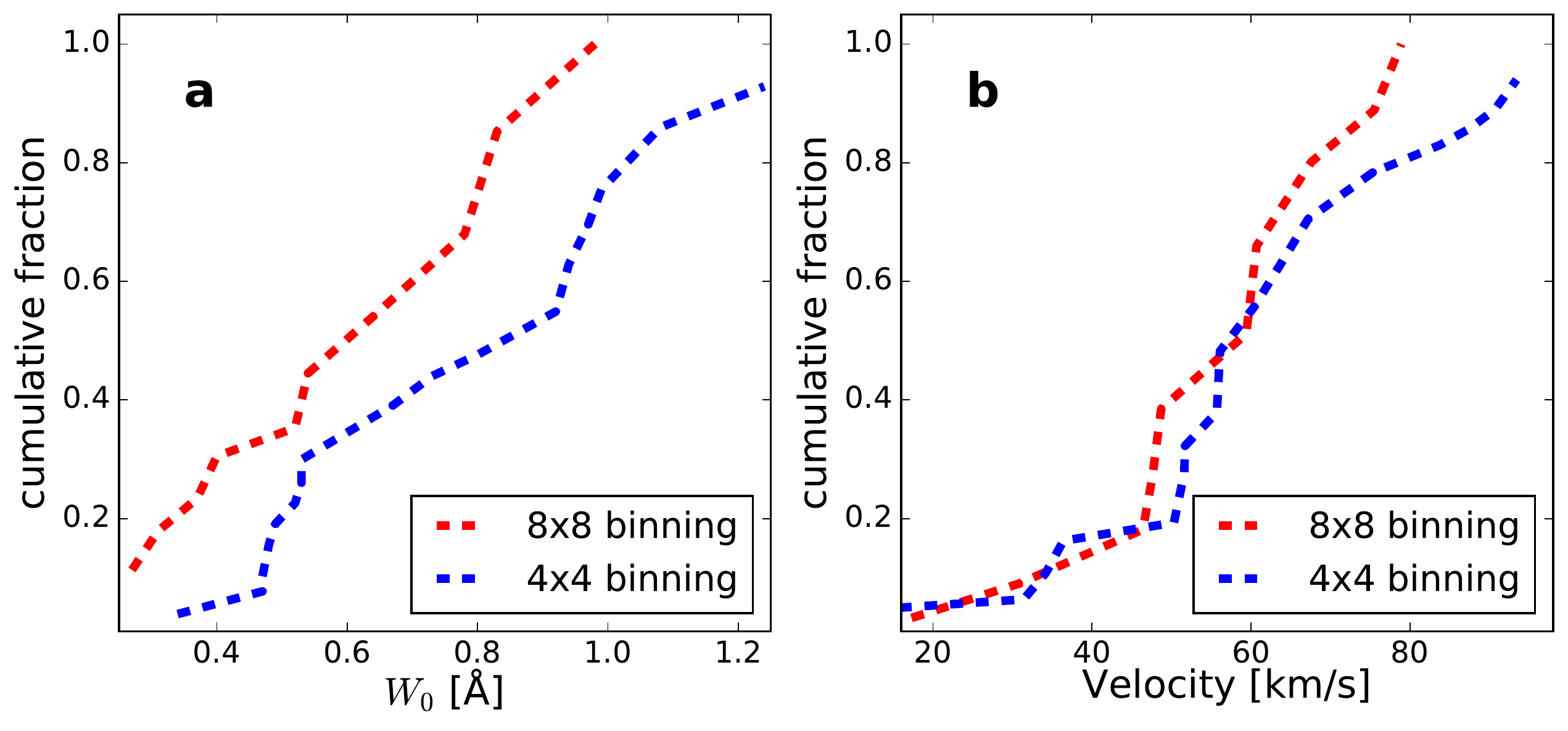}
\caption{{\bf Effect of  partial covering.}  
{\bf (a):} Cumulative distribution of absorption-strengths for two different
binnings. {\bf (b): } Same for velocities. 
\label{fig_covering}
}
\end{figure}

\renewcommand{\tablename}{Extended Data Table}
\setcounter{figure}{0}

\clearpage

\begin{table}
\scriptsize
\centering
 \begin{minipage}{90mm}
\caption{ \mgii\ absorption near G1}
\medskip
\begin{tabular}{RRCCCC}
\hline
\hline
\Delta\alpha^a & \Delta\delta^a & D^b & W_0^{{ } c} & v^d & {\rm S/N}^e\\
(\arcsec) & (\arcsec) & [{\rm kpc}] & [{\rm \AA}] & [{\rm km~s}^{-1}] & \\
\hline
\hline
-22.0&-11.2&77.5&<0.76&...&3.6\\
4.4&-10.4&40.3&<0.76&...&3.6\\
3.6&-10.4&40.5&0.53\pm0.14&55.7\pm13.8&8.7\\
2.8&-10.4&41.3&<0.59&...&4.6\\
-21.2&-10.4&76.5&<0.60&...&4.5\\
-22.0&-10.4&78.0&<0.79&...&3.4\\
4.4&-9.6&38.3&0.52\pm0.18&15.7\pm26.3&6.8\\
3.6&-9.6&38.5&<0.26&...&10.6\\
2.8&-9.6&39.2&<0.82&...&3.3\\
-21.2&-9.6&77.2&<0.33&...&8.1\\
-22.0&-9.6&78.4&<0.69&...&3.9\\
4.4&-8.8&36.2&<0.49&...&5.5\\
3.6&-8.8&36.2&0.34\pm0.12&50.3\pm15.8&8.7\\
-21.2&-8.8&77.7&<0.29&...&9.2\\
-22.0&-8.8&79.5&<0.28&...&9.8\\
-22.8&-8.8&81.1&<0.88&...&3.1\\
3.6&-8.0&33.8&0.73\pm0.24&60.4\pm25.0&5.1\\
2.8&-8.0&34.3&<0.50&...&5.4\\
-21.2&-8.0&77.8&<0.48&...&5.6\\
-22.0&-8.0&79.9&<0.32&...&8.5\\
-22.8&-8.0&81.8&<0.38&...&7.1\\
3.6&-7.2&31.2&<0.79&...&3.5\\
2.8&-7.2&31.5&<0.24&...&11.3\\
2.0&-7.2&32.6&<0.40&...&6.8\\
-22.0&-7.2&80.1&<0.65&...&4.2\\
-22.8&-7.2&82.4&<0.45&...&6.1\\
2.8&-6.4&28.6&0.48\pm0.17&34.2\pm23.5&9.8\\
2.0&-6.4&29.5&<0.23&...&12.1\\
1.2&-6.4&30.7&<0.56&...&4.8\\
-22.8&-6.4&82.8&<0.90&...&3.0\\
2.0&-5.6&26.3&0.47\pm0.14&93.4\pm25.9&7.4\\
1.2&-5.6&27.4&0.47\pm0.10&51.7\pm18.0&10.6\\
0.4&-5.6&28.9&<0.56&...&4.9\\
1.2&-4.8&23.8&<0.54&...&5.1\\
0.4&-4.8&25.2&0.53\pm0.12&63.5\pm12.3&7.2\\
-0.4&-4.8&27.1&1.24\pm0.43&115.4\pm29.0&3.2\\
0.4&-4.0&21.4&<0.85&...&3.2\\
-0.4&-4.0&23.2&<0.59&...&4.6\\
-1.2&-4.0&25.4&0.67\pm0.26&83.7\pm16.1&3.2\\
-1.2&-3.2&21.3&0.49\pm0.11&75.3\pm9.9&7.9\\
-2.0&-3.2&23.8&0.99\pm0.15&90.7\pm9.7&8.3\\
-1.2&-2.4&17.1&1.08\pm0.19&87.5\pm12.6&7.8\\
-2.0&-2.4&19.7&0.92\pm0.11&51.6\pm7.3&11.0\\
-2.8&-2.4&22.4&0.8\pm0.24&31.4\pm19.0&5.2\\
-3.6&-1.6&21.7&1.42\pm0.20&36.4\pm11.3&5.0\\
-4.4&-1.6&24.7&0.94\pm0.14&56.1\pm9.2&9.0\\
-5.2&-1.6&27.3&<0.61&...&4.5\\
-9.2&-1.6&40.6&<0.67&...&4.1\\
-5.2&-0.8&24.5&<0.63&...&4.3\\
-6.0&-0.8&27.6&0.97\pm0.26&67.2\pm17.5&3.4\\

\hline
\label{table_absorption}
\end{tabular}

{\fontfamily{ptm}\selectfont
$^a${\footnotesize Arc-position angular separation to G1 in the image plane};
$^b${\footnotesize Projected physical separation to G1 in the absorber plane};
$^c${\footnotesize  \mgii$\lambda 2796$ absorption-strength (with $1\sigma$
  error) or $2\sigma$ upper limit;} 
$^d${\footnotesize   Velocity relative to \zg$ = 0.98235$;}
$^e${\footnotesize Signal-to-noise ratio to the blue of  \mgii }
}
\end{minipage}
\end{table}

\clearpage

\begin{table}
\footnotesize
\centering
 \begin{minipage}{80mm}
\caption{ Upper limits on \mgii\ absorption near G2}
\medskip
\begin{tabular}{RRCCCC}
\hline
\hline
\Delta\alpha^a & \Delta\delta^a & D^b & W_0^c & v^d & {\rm S/N}^e\\
(\arcsec) & (\arcsec) & [{\rm kpc}] & [{\rm \AA}] & [{\rm km~s}^{-1}] & \\
\hline
\hline
-4.2&-1.6&22.0&<0.81&...&3.4\\
-5.0&-1.6&25.2&<0.70&...&3.9\\
-5.8&-1.6&28.9&<0.32&...&8.6\\
-6.6&-1.6&32.5&<0.41&...&6.7\\
-5.8&-0.8&27.1&<0.79&...&3.4\\
-6.6&-0.8&30.9&<0.60&...&4.5\\

\hline
\label{table_absorption_G2}
\end{tabular}

{\fontfamily{ptm}\selectfont
$^a${\footnotesize Arc-position angular separation to G2 in the image plane};
$^b${\footnotesize Projected physical separation to G2 in the absorber plane};
$^c${\footnotesize  \mgii$\lambda 2796$ absorption-strength $2\sigma$
    upper limit;}  
$^d${\footnotesize   Velocity relative to \zg$ = 0.98235$;}
$^e${\footnotesize Signal-to-noise ratio to the blue of  \mgii }
}
\end{minipage}
\end{table}

\clearpage

\begin{table}
\footnotesize
\centering
 \begin{minipage}{100mm}
\caption{ Absorption by \feii\ and \mgi\ near G1}
\medskip
\begin{tabular}{RRCCCC}
\hline
\hline
\Delta\alpha & \Delta\delta & D & W_0^{2796}  & W_0^{2382}  &  W_0^{2852}\\
(\arcsec) & (\arcsec) & [{\rm kpc}] & [{\rm \AA}] & [{\rm \AA}]   & [{\rm \AA}] \\
\hline
\hline
-2.0&-3.2&23.8&0.99\pm0.15&0.72\pm0.22&<0.35\\
-1.2&-2.4&17.1&1.08\pm0.19&0.59\pm0.21&0.44\pm0.15\\
-4.4&-1.6&24.7&0.94\pm0.14&0.62\pm0.19&<0.31\\
\hline
\label{table_absorption_FeII}
\end{tabular}

\end{minipage}
\end{table}

\clearpage

\begin{table}
\footnotesize
\centering
\caption{ Galaxy properties}
\medskip
\begin{tabular}{LCCCCCCC}
\hline
{\rm ID}& {\rm RA} & {\rm DEC} & z & v   &\Delta v_{\rm FWHM} & m_{\rm  F814W} &B-I\\
  & [{\rm deg}] & [{\rm deg}] & &[{\rm km~s}^{-1}]   &[{\rm km~s}^{-1}]    & {\rm  (AB)} & \\
(1) & (2) &(3)&(4)&(5)&(6)&(7)&(8)\\
%\hline
\hline
G1{\rm -}A  &51.867229&-13.434300&0.98236 & +1.8\pm7.0    & 138.0\pm 10.4&24.30\pm0.03&0.68\pm0.03\\
G1{\rm -}B  &51.867420&-13.434390&0.98267 & +48.8\pm12.0  & 97.9\pm 18.4 &24.64\pm0.04&0.53\pm0.04\\
G1{\rm -}C  &51.867229&-13.434060&0.98212 & -35.2\pm12.5 & 92.5\pm 17.9 &24.85\pm0.05&0.72\pm0.05\\
G2          &51.865139&-13.447130&0.98216 & -29.2\pm1.7  & 81.0\pm 2.8  &24.27\pm0.03&0.06\pm0.03\\
G3          &51.855511&-13.431970&0.98162 & -109.8\pm12.2& 86.9\pm 17.4 &23.99\pm0.02&0.30\pm0.02\\

\hline
{\rm ID}& M_B& L/L^* & \log{M_*/{\rm M}_{\rm   \astrosun}} & f_{\rm OII}
& {\rm SFR}&\mu &R_{\rm vir}\\
        &    &      &               &[{\rm erg~cm}^{-2}~{\rm s}^{-1}] &[{\rm
    M}_{\rm   \astrosun}~{\rm  yr}^{-1}]&&{\rm [kpc]}\\
 &(9) & (10) &(11)&(12)&(13)&(14)&(15)\\
\hline
G1{\rm -}A  &-18.37&  0.05  &  9.5&8.7\cdot10^{-18}&0.23&2.6&92\\
G1{\rm -}B  &-17.92&  0.03  &  9.1&4.0\cdot10^{-18}&0.11&2.7&79\\
G1{\rm -}C  &-17.81&  0.03  &  9.1&3.3\cdot10^{-18}&0.09&2.5&79\\
G2          &-18.86&  0.08  &  9.0&2.4\cdot10^{-17}&0.85&2.0&73\\
G3          &-19.08&  0.10  &  9.4&6.9\cdot10^{-18}&0.28&1.7&85\\

\hline
\label{table_galaxies}
\end{tabular}
{\fontfamily{ptm}\selectfont
{ 
Columns.---
(1)--(3) Galaxy identification and coordinates:
(4) redshift;
(5) velocity relative to \zg=$0.98235$;
(6) deconvolved [\oii] line width;
(7) apparent magnitude;
(8) rest-frame color;
(9) absolute magnitude;
(10) rest-frame $B$-band luminosity  in terms of $L^*$ at
$z=0.98$\cite{Willmer2006};   
(11) stellar mass (from SED fitting);
(12) [\oii] emission line flux
(13) star formation rate (from emission line flux);
(14) magnification (subject to $\approx 5$\%\ uncertainties). Magnification
was  used to correct quantities (9)--(13);
(15) virial radius. 
}
}
\end{table}

\setcounter{figure}{0}

\end{document}